\documentclass[final,3p,times,onecolumn]{ms}

\usepackage{amssymb,amsfonts,amsmath,amstext,amsgen,amsopn,amsxtra,indentfirst,lscape,rotating,epsfig}

\journal{Kevin France (Kevin.France@colorado.edu)}

\begin{document}

\begin{frontmatter}

\title{Observing Supernova 1987A with the Refurbished Hubble Space Telescope}

\author[casa]{Kevin France$^{*,}$}
\address[casa]{Center for Astrophysics and Space Astronomy, University of Colorado, Boulder, CO 80309-0389, U.S.A.}
 
\author[jila]{Richard McCray}
\address[jila]{JILA, University of Colorado, Boulder, CO 80309-0440, U.S.A.}

\author[eth,ias]{Kevin Heng}
\address[eth]{ETH Z\"{u}rich, Institute for Astronomy, Wolfgang-Pauli-Strasse 27, CH-8093, Z\"{u}rich, Switzerland}
\address[ias]{Institute for Advanced Study, School of Natural Sciences, Einstein Drive, Princeton, NJ 08540, U.S.A.}

\author[cfa]{Robert P. Kirshner}
\address[cfa]{Harvard-Smithsonian Center for Astrophysics, 60 Garden Street, Cambridge, MA 02138, U.S.A.}

\author[cfa]{Peter Challis}

\author[cea]{Patrice Bouchet}
\address[cea]{Service d'Astrophysique DSM/IRFU/SAp CEA - Saclay, Orme des Merisiers, FR 91191 Gif-sur-Yvette,
France}

\author[columbia]{Arlin Crotts}
\address[columbia]{Department of Astronomy, Mail Code 5240, Columbia University, 550 W. 120th St., New York, NY 10027, U.S.A.}

\author[gsfc]{Eli Dwek}
\address[gsfc]{NASA Goddard Space Flight Center, Code 665, Greenbelt, MD 20771, U.S.A.}

\author[stockholm]{Claes Fransson}
\address[stockholm]{Department of Astronomy, The Oskar Klein Centre, Stockholm University, 106 91 Stockholm, Sweden}

\author[nd]{Peter M. Garnavich}
\address[nd]{225 Nieuwland Science, University of Notre Dame, Notre Dame, IN 46556-5670, U.S.A.}

\author[stockholm]{Josefin Larsson}

\author[hofstra]{Stephen S. Lawrence}
\address[hofstra]{Department of Physics and Astronomy, Hofstra University, Hempstead, NY 11549, U.S.A.}

\author[stockholm]{Peter Lundqvist}

\author[stsci,inaf,snltd]{Nino Panagia}
\address[stsci]{ Space Telescope Science Institute, 3700 San Martin Drive, Baltimore, MD 21218, U.S.A.}
\address[inaf]{ INAF/CT, Osservatorio Astrofisico di Catania, Via S. Sofia 78, I-95123 Catania, Italy}
\address[snltd]{ Supernova Ltd, OYV \#131, Northsound Road, Virgin Gorda, British Virgin Islands}

\author[hongkong]{Chun S. J. Pun}
\address[hongkong]{Department of Physics, The University of Hong Kong,
Pok Fu Lam Road, Hong Kong, China}

\author[ucb]{Nathan Smith}
\address[ucb]{Department of Astronomy, University of California, Berkeley, CA 94720-3411, U.S.A.}

\author[stockholm]{Jesper Sollerman}

\author[gsfc]{George Sonneborn}

\author[casa]{John T. Stocke}

\author[texas_am]{Lifan Wang}
\address[texas_am]{Department of Physics \& Astronomy, Texas A\&M University, College Station, TX 77843-4242, U.S.A.}

\author[texas]{J. Craig Wheeler}
\address[texas]{Department of Astronomy, University of Texas, Austin, TX 78712-0259, U.S.A.}


\begin{abstract} 
Observations with the Hubble Space Telescope (HST), conducted since 1990,  
now offer an unprecedented glimpse into fast astrophysical shocks in the young  
remnant of supernova 1987A.  Comparing observations taken in 2010  
using the refurbished instruments on HST with data taken in 2004, just  
before the Space Telescope Imaging Spectrograph failed, we find that the Ly$\alpha$ and H$\alpha$ lines from shock emission continue to brighten, while their maximum velocities continue to decrease. We observe broad blueshifted Ly$\alpha$, which we attribute to resonant scattering of photons emitted from hotspots on the equatorial ring.  We also detect N~{\sc v}~$\lambda\lambda$1239,1243 \AA~ line emission, but only to the red of Ly$\alpha$. The profiles of the N~{\sc v} lines differ markedly from that of H$\alpha$, suggesting that the N$^{4+}$ ions are scattered and accelerated by turbulent electromagnetic fields that isotropize the ions in the collisionless shock. 

\end{abstract}


\end{frontmatter} 

The death of a massive star produces a violent explosion known as a supernova (SN), which expels matter at hypersonic velocities.  Supernovae deposit large amounts of mechanical energy and nucleosynthesized elements into the surrounding interstellar medium, driving the physical and chemical evolution of galaxies.
The shock impact of the supernova debris with ambient matter creates a radiating system known as a supernova remnant.  SN 1987A (23 February 1987), the brightest such event observed since Kepler's supernova (SN~1604) \citep{arnett89}, provides a unique opportunity to witness the development of a supernova remnant \citep{mccray93,mccray07}.  
Because SN 1987A is only 50 kpc away in the Large Magellanic Cloud,  
HST's superb angular resolution is sufficient to resolve the  
interaction of its shocks with circumstellar material.
To track and interpret the temporal, spatial, and spectral evolution of SN1987A, we present observations obtained on 31 January 2010 with the recently repaired Space Telescope Imaging Spectrograph ({\it STIS}) and compare them with the results from the last epoch of observations (18-23 July 2004) \citep{heng06}, prior to the instrument's failure in August 2004.   

The rapidly expanding debris of a supernova explosion interacts hydrodynamically with circumstellar matter.  If the circumstellar matter has a smooth density distribution, a double-shock structure will be established \citep{chevalier82}. A forward shock (blast wave) propagates into the circumstellar matter, creating a layer of hot, shocked gas. The pressure of this layer drives a reverse shock into the supernova debris. This double-shock structure propagates outwards until the blast wave encounters a relatively dense obstacle. 
In the case of SN 1987A, encountering the equatorial ring drives a reflected shock backwards into the debris that merges it with the reverse shock.  

The equatorial ring is a relatively dense ($n_{\rm H} \sim 10^3$--$10^4$ atoms cm$^{-3}$)  structure of diameter 1.34 light years (ly), inclined at an angle $i = 45^\circ$ with respect to the line of sight \citep{panagia91,lundqvist96,groningsson08}.  This ring is attributed to a mass loss event that occurred about 20,000 years before the supernova explosion \citep{crotts91,mp07}.  
The first evidence of interaction of the blast wave with the ring appeared in 1995, when a rapidly brightening optical ``hotspot'' appeared in images taken with the Wide Field Planetary Camera 2 ({\it WFPC2}) aboard the HST \citep{garnavich97,lawrence00}.  Today, the ring is encircled by about 30 hotspots (Fig. 1). The associated on-line movie shows the emergence of these hotspots and the expansion of the debris over the past 15 years.  The location of these emission spots just inside the ring and their  
long duration suggests they result from shocks that propagate into  
dense fingers that protrude inwards from the equatorial ring~\citep{pun02}. 

\begin{figure}[t]
\vspace{-1.5in}
\begin{center}
\hspace{+0.0in}
\epsfig{figure=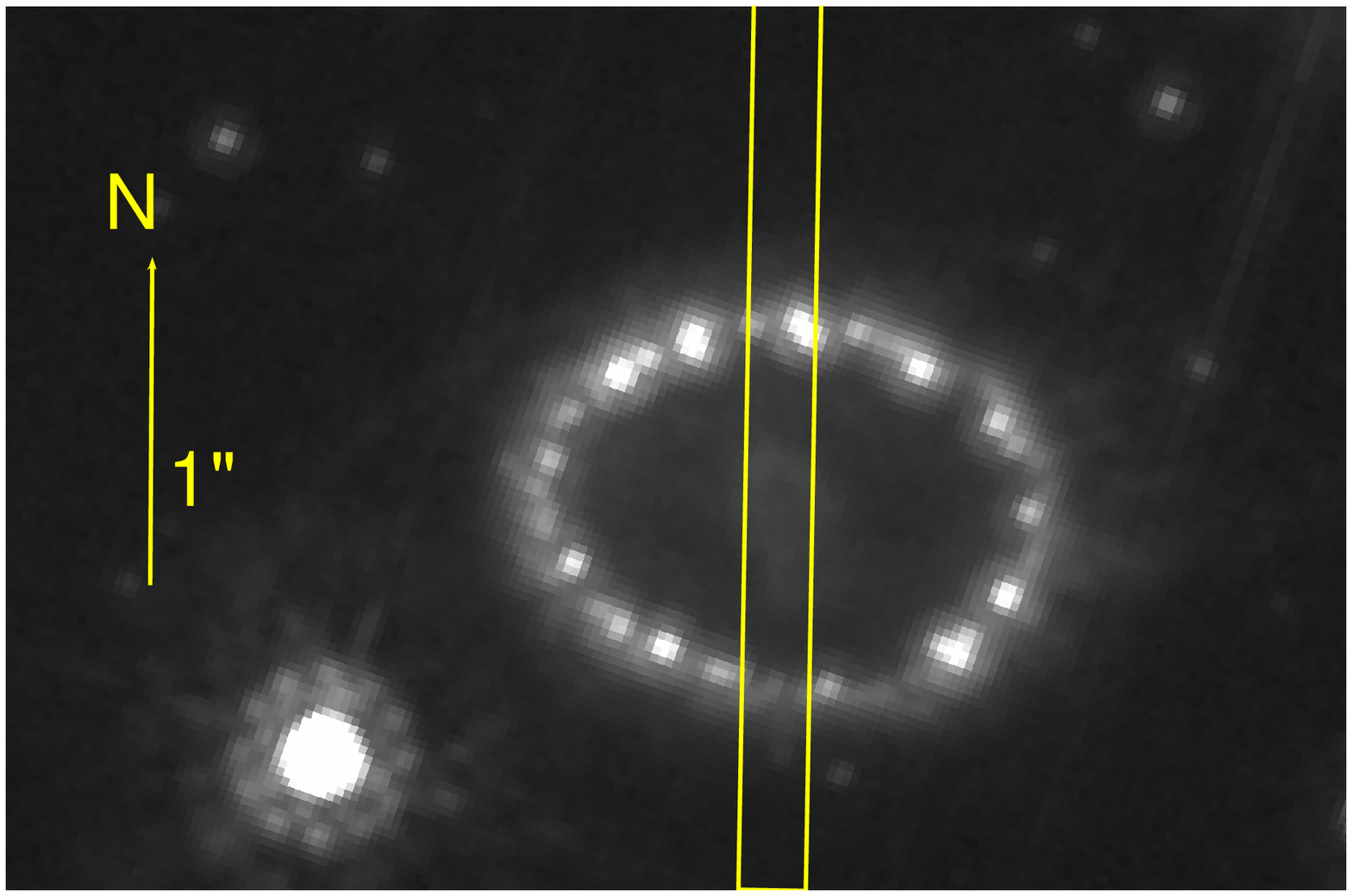,width=2.5in,angle=00}
\hspace{0.3in}
\epsfig{figure=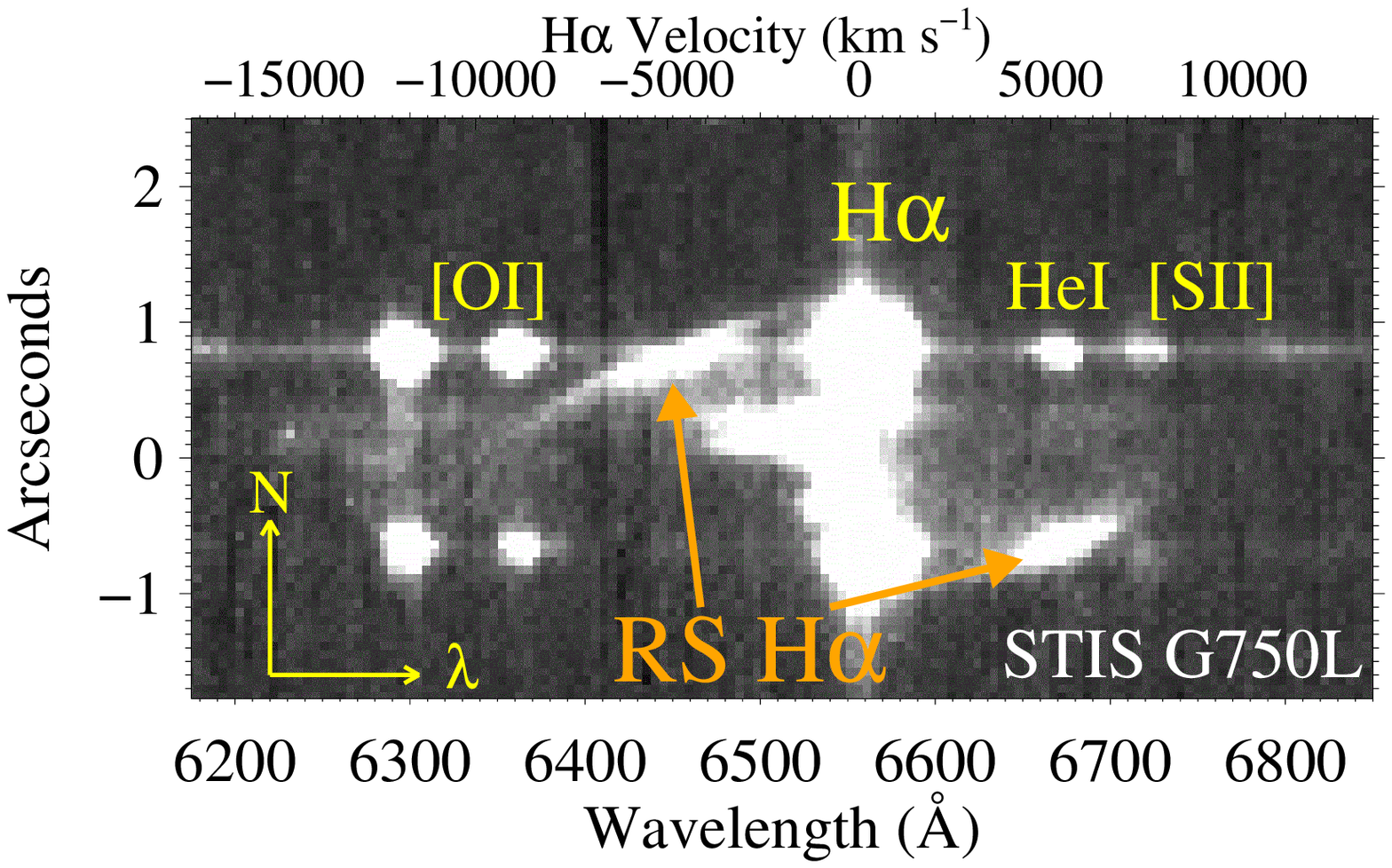,width=3.2in,angle=90}
\caption{({\it left}) HST-Advanced Camera for Surveys image obtained on 28 November 2003 in the F625W filter with an exposure time of 800s, illustrating the slit orientation used. 
({\it right}) {\it STIS} G750L spectrum of SN 1987A obtained on 31 January 2010, centered about the H$\alpha$ emission line.  
The vertical bar at the center of the image is stationary H$\alpha$ emission from interstellar or circumstellar gas. The bright spots at the north and south of this bar 
are the emissions from H$\alpha$~+~[N~{\sc ii}] $\lambda\lambda$6548,6583~\AA\ from hotspots on the equatorial ring.  [O~{\sc i}] $\lambda\lambda$6300,6364, He~{\sc i}~$\lambda$6678, and [S~{\sc ii}]~$\lambda\lambda$6716,6731 \AA\ emission from the hotspots is also observed.  The blueshifted streaks near the center are H$\alpha$ emission excited by radioactivity in the interior of the supernova debris.  The curved, blueshifted streak extending from the north side of the vertical bar and the redshifted streak on the south side (noted with orange arrows) are H$\alpha$ emission from the brightest parts of the reverse shock (RS).} 
\end{center}
\end{figure}

Radiation from the reverse shock can be observed at optical and ultraviolet wavelengths.  Before it reaches the reverse shock, the outer layer of the supernova debris consists mostly of partially ionized hydrogen and helium gas that has been expanding freely since the explosion.  When neutral hydrogen atoms cross the reverse shock, they are excited and ionized by collisions with ions in the shocked plasma. If the atoms are excited before they are ionized, they will produce Ly$\alpha$ (1216 \AA) and H$\alpha$ (6563 \AA) line emission. On average, approximately 1 Ly$\alpha$ photon and 0.2 H$\alpha$ photons are produced for every hydrogen atom crossing the reverse shock \citep{CKR80,michael03,heng07}.

The emission properties of the reverse shock in SN 1987A are similar to those of the Balmer-dominated shock emission observed in several supernova remnants, where photons are produced via collisional excitation (and charge exchange) rather than recombination \citep{CKR80,cr78,heng10}.  The difference is that, in other collisionless supernova remnants the  
blast wave overtakes nearly stationary circumstellar matter, while in  
SN 1987A fast-moving hydrogen gas in the supernova debris overtakes the reverse shock. 
As the hydrogen atoms in the supernova debris cross the reverse shock, they freely stream with radial velocity $v_{r}$ = $r$/$t_{e}$, where $r$ is the radius of the reverse shock measured from the explosion center and $t_{e}$ is the time since the explosion. Likewise, the atoms have Doppler velocity (projected along the line of sight) $v_{z} = -r\cos \theta/t_{e}$, where $\theta$ is the angle between the streaming supernova debris and the line of sight to the observer. When they are excited by collisions with the shocked gas, the neutral hydrogen atoms are not deflected, so the Doppler shifts of the resulting emission lines we observe (Fig. 1) correspond to the projected ballistic velocity of the unshocked supernova debris.

\begin{figure}[h]
\begin{center}
\hspace{+0.0in}
\epsfig{figure=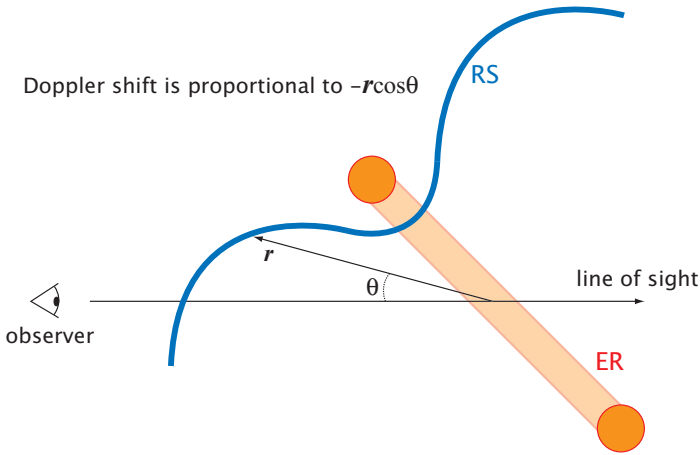,width=3.75in,angle=00}
\epsfig{figure=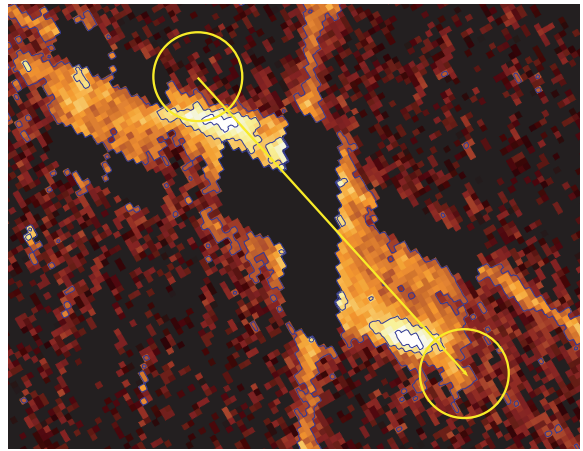,width=3.0in,angle=00}
\caption{({\it left}) Schematic illustration of the location of the reverse shock (RS) with respect to the equatorial ring (ER), which is inclined $45^\circ$ to the line of sight. ({\it right})  H$\alpha$ emission from the reverse shock transformed into a cross-sectional view through the circumstellar ring (see text). The diameter of the equatorial ring is 1.34 ly.  The left and right yellow circles represent the near (N) and far (S) sides of the equatorial plane of the ring, respectively.  
The circles represent the cross section of the ring, and contours highlight the emission peaks on the northern and southern streaks.}
\end{center}
\end{figure}

The unique mapping between distance along the line-of-sight and  
Doppler shift allows us to convert the spectrum of Figure 1 into a map  
of the location of the reverse shock (Fig. 2).  The H$\alpha$ emission is  
concentrated just inside the equatorial ring because at these  
locations the reverse shock penetrates into denser regions of the  
supernova envelope, where the flow is held back by shocks reflected  
from the ring. 
The net H$\alpha$ flux observed through the $52^{\prime\prime} \times 0.2^{\prime\prime}$ slit in the total reverse shock (northern blueshifted plus southern redshifted streaks, Fig. 1) is 3.3~($\pm$~0.5)~$\times$~10$^{-13}$ erg cm$^{-2}$ s$^{-1}$~\citep{n1}.  This value is a factor of about 1.7 greater than that measured in July 2004 \citep{heng06} and February 2005~\citep{smith05}.

\begin{figure}[t]
\vspace{-1.7in}
\begin{center}
\hspace{+0.0in}
\epsfig{figure=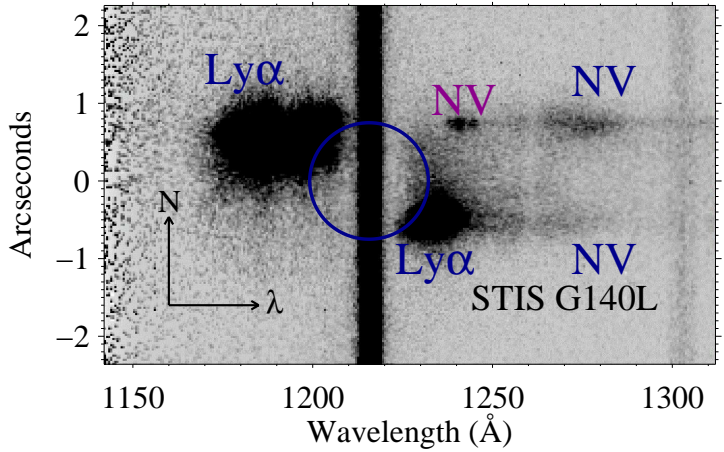,width=4.0in,angle=00}
\caption{{\it STIS} G140L observations of SN 1987A, acquired on 31 January 2010.   The bright vertical stripe is the slit image in geocoronal Ly$\alpha$.  The blue ellipse approximates the location of the circumstellar emission ring. Broad, faint features seen on the north and south sides between $\sim 1260$--1290 \AA~ (labeled N~{\sc v} in blue) are produced by nitrogen ions in the reverse shock, and the shock-excited N~{\sc v} hotspot emission is labeled in magenta.
}  
\end{center}
\end{figure}

Unlike H$\alpha$ photons, Ly$\alpha$ photons experience resonant scattering by hydrogen atoms.  Thus, Ly$\alpha$ emission is not observed at wavelengths immediately blue and redward of 1216~\AA\ because of scattering by hydrogen atoms in the Milky Way and Large Magellanic Cloud (Fig. 3).  
Figure 4 shows comparisons of one-dimensional scans of the Ly$\alpha$ (the dark streaks in Figure 3) and H$\alpha$ (the bright streaks in Figure 1) emission from the reverse shock. The observed H$\alpha$ and Ly$\alpha$ fluxes have been increased by factors $\approx$~1.5 and 8, respectively, to correct for interstellar extinction along the line of sight to SN 1987A \citep{suntzeff90,scuderi96}.  
On the north side of the equatorial ring, the ratio of Ly$\alpha$ to H$\alpha$ photon fluxes has a fairly constant value near 40 for velocities between $-2500$ km~s$^{-1}$ and $-6000$ km~s$^{-1}$ (Fig. 4).  This ratio is much greater than the expected photon production ratio of 5:1 for a Balmer-dominated shock \citep{michael03,heng07}.  Moreover, the H$\alpha$ emission fades for blueshift velocities $<$~$-$8,000 km s$^{-1}$, while the Ly$\alpha$ emission remains bright to blueshift velocities approaching $-$12,000 km s$^{-1}$.	
If the Ly$\alpha$ photons are produced by the same mechanism as the H$\alpha$ photons, then the Ly$\alpha$ to H$\alpha$ ratio should be the same for all observed velocities; but it is not.  Therefore, most of the observed Ly$\alpha$ emission cannot be produced directly by hydrogen atoms crossing the reverse shock.  Unlike H$\alpha$, the broad Ly$\alpha$ emission is not confined to a narrow strip delineating the reverse shock surface (Fig. 3).

\begin{figure}
\begin{center}
\hspace{+0.0in}
\epsfig{figure=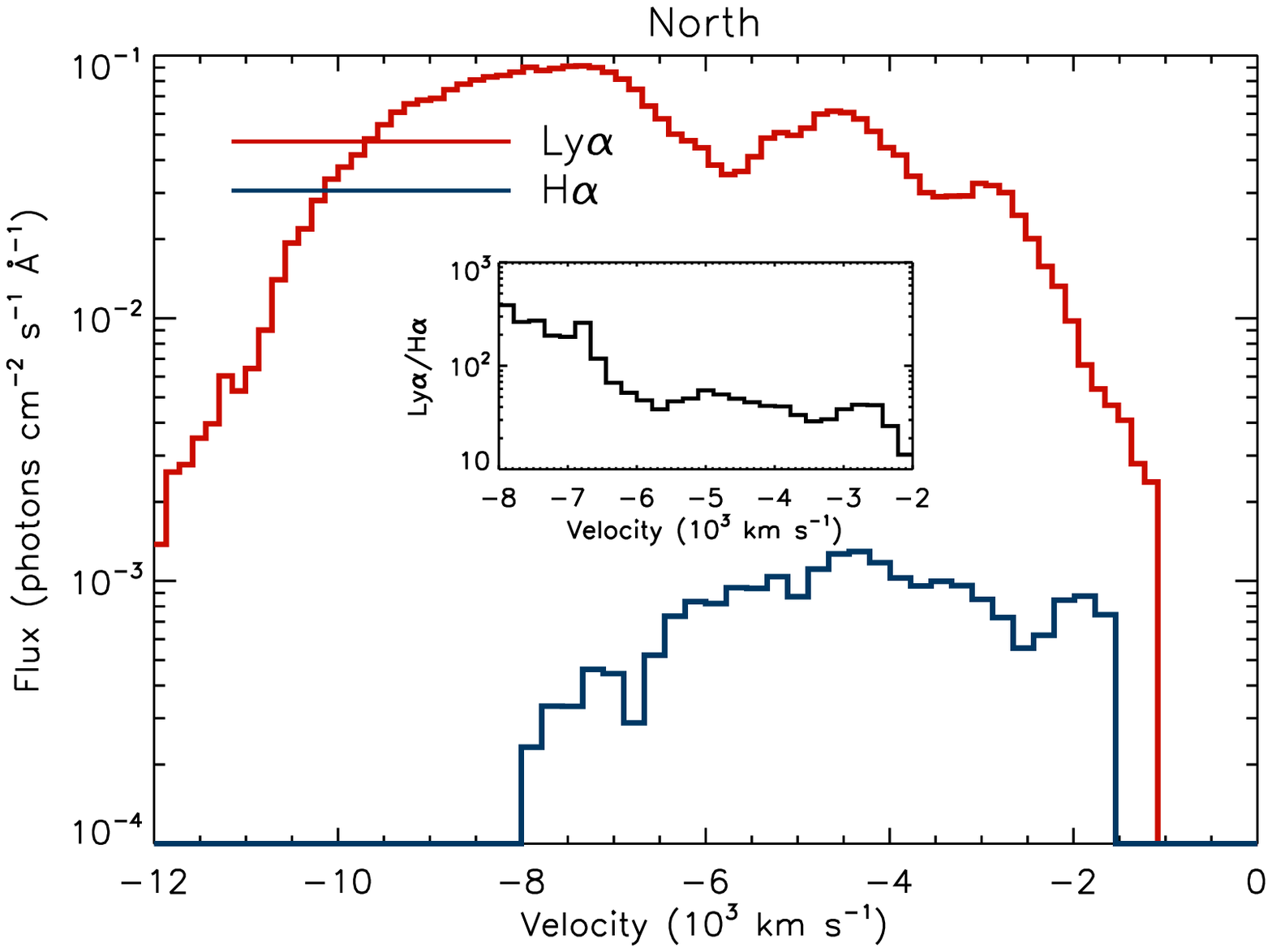,width=3.3in,angle=0}
\epsfig{figure=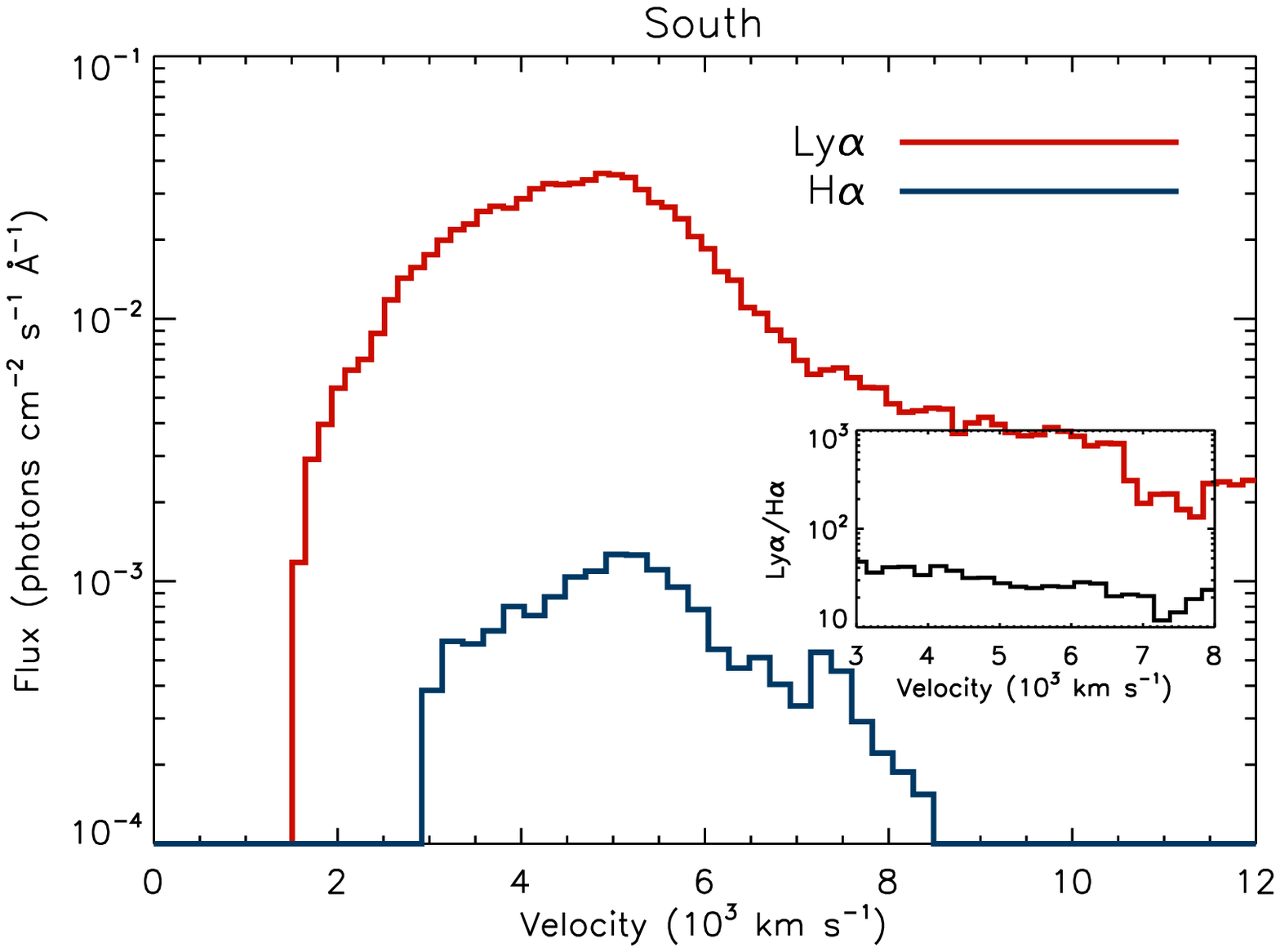,width=3.3in,angle=0}
\caption{({\it left}) Integrated H$\alpha$ and Ly$\alpha$ emission from the reverse shock on the north side of the equatorial ring.  While the H$\alpha$ emission approaches zero blueward of $-8000$ km s$^{-1}$, it has been artificially truncated in this plot because of strong contamination from [O~{\sc i}] $\lambda\lambda$6300,6364 \AA~ present in the hotspot (Fig. 1).  The inset shows the ratio of Ly$\alpha$ to H$\alpha$. The broad dip in the Ly$\alpha$ emission near $-6000$ km s$^{-1}$ may be due to absorption by interstellar Si~{\sc ii} ($\lambda\lambda$1190,1193~\AA).  Interstellar absorptions from Si~{\sc ii} ($\lambda\lambda$1190,1193 \AA, $\lambda\lambda$1260,1265 \AA~ and $\lambda\lambda$1526,1533 \AA) are clearly detected on the sightline to a nearby star.
({\it right}) Velocity profiles for the southern, redshifted emission.  We truncate the southern H$\alpha$ profile at $v \ge 8500$ km s$^{-1}$ because the H$\alpha$ flux at these velocities is consistent with a combination of detector background and [S~{\sc ii}] emission scattered inside of the equatorial ring radius.}
\end{center}
\end{figure}

We propose a different mechanism to account for the highly blueshifted Ly$\alpha$ emission.  As the supernova blast wave enters the equatorial ring, the shocked hotspots on the ring become bright sources of Ly$\alpha$ radiation.  This radiation is invisible to observers on Earth because it is centered at zero velocity with respect to the interstellar neutral hydrogen and its linewidth is narrow ($\Delta v < 300$ km s$^{-1}$) \citep{groningsson08,pun02}, so that it is entirely blocked by the interstellar medium.  Roughly half of this radiation propagates inwards into the supernova debris, where the Ly$\alpha$ photons may be resonantly scattered by hydrogen atoms that are expanding with radial velocities ranging from 3000 km s$^{-1}$ to 9000 km s$^{-1}$.     In the rest frame of the hydrogen atoms in the expanding debris, photons propagating inwards are blueshifted.  If they are then scattered backwards towards Earth, they will be blueshifted a second time. 
There is no corresponding bright, high-velocity, redshifted Ly$\alpha$ component on the south side of the equatorial ring (Figs 3, 4).  The Ly$\alpha$ photons which are emitted radially inwards by the ring hotspots on the south side are seen as blueshifted by hydrogen atoms in the onrushing debris. Unlike the case in the north, photons from the south that are scattered towards the observer receive a redshift that cancels out this blueshift.  This leaves the Ly$\alpha$ photons at approximately zero-velocity, where they will be scattered or absorbed before reaching us.  

To check the plausibility of such a mechanism, it is necessary to verify that a sufficient number of Ly$\alpha$ photons are emitted by the hotspots to account for the observed high-velocity Ly$\alpha$ and that the neutral hydrogen layer in the expanding supernova envelope has sufficient optical depth to scatter Ly$\alpha$ photons by roughly half of the observed maximum velocity, {\it i.e.}, 6000 km s$^{-1}$.  It is not possible to measure the emitted Ly$\alpha$ flux from the hotspots directly because this radiation is blocked by interstellar hydrogen atoms. We can estimate the strength of Ly$\alpha$ emission using 
the direct H$\alpha$ emission from the ring,  
applying a theoretical scaling (Ly$\alpha$:H$\alpha$~$\sim 5-10$; \citep{osterbrock89}).  The H$\alpha$ hotspot emission within the {\it STIS} slit in the 2010 observations ($\approx$~2.1$\times$~10$^{-12}$ erg cm$^{-2}$ s$^{-1}$) suggests that the hotspot Ly$\alpha$ flux in 2010 is approximately 3--7 photons cm$^{-2}$ s$^{-1}$. This is sufficient to account for the broad blueshifted Ly$\alpha$ emission (Fig. 4). 

Regarding the question of optical depth in the line wings of Ly$\alpha$, the radial column density of hydrogen atoms in the debris is roughly $N_{\rm H} \approx M_{\rm H}/(4\pi R^2 m_{\rm H})$, where $M_{\rm H} \sim 5 M_\odot$ is the estimated mass of hydrogen in the supernova envelope, $R \approx 6 \times 10^{17}$ cm is the radius of the envelope, and $m_{\rm H}$ is the mass of the hydrogen atom.  For these estimated values, $N_{\rm H} \approx 1.3 \times 10^{21}$ cm$^{-2}$.  The optical depth of such a column to Ly$\alpha$ photons, Doppler shifted by $v_{1000} = v/(1000$ km s$^{-1})$, is given roughly by $\tau$~$\approx$~$N_{\rm H}$~$\sigma_{o}$( ($\gamma$/4$\pi^{2}$) /($\Delta f^{2}$ + ($\gamma$/4$\pi$)$^{2}$) ) \citep{rybicki79}, where the frequency shift is $\Delta f/f_0 = v/c$, $\sigma_{o}$ is the line center absorption cross section, and $\gamma = 6.3 \times 10^{8}$~s$^{-1}$ is the spontaneous decay rate of electrons in the excited state of the hydrogen atom leading to the emission of Ly$\alpha$ photons.  For $v_{1000}$~=~$-$6, $\tau$~$\approx$~0.1;  for comparison, $\tau$~$>$~2.5 for $|v_{1000}|$~$<$~1.  
Our estimate of the available Ly$\alpha$ photon budget from the hotspots provides ample margin for this mechanism to operate,  
even if $<$~50\% of the hotspot Ly$\alpha$ enters the debris.
A measurable fraction of the Ly$\alpha$ photons emitted by a hotspot that enter the supernova debris will be backscattered and emerge with blueshifts ranging up to $\sim$~$-$12,000 km s$^{-1}$.

\begin{figure}
\begin{center}
\vspace{-1in}
\hspace{+0.0in}
\epsfig{figure=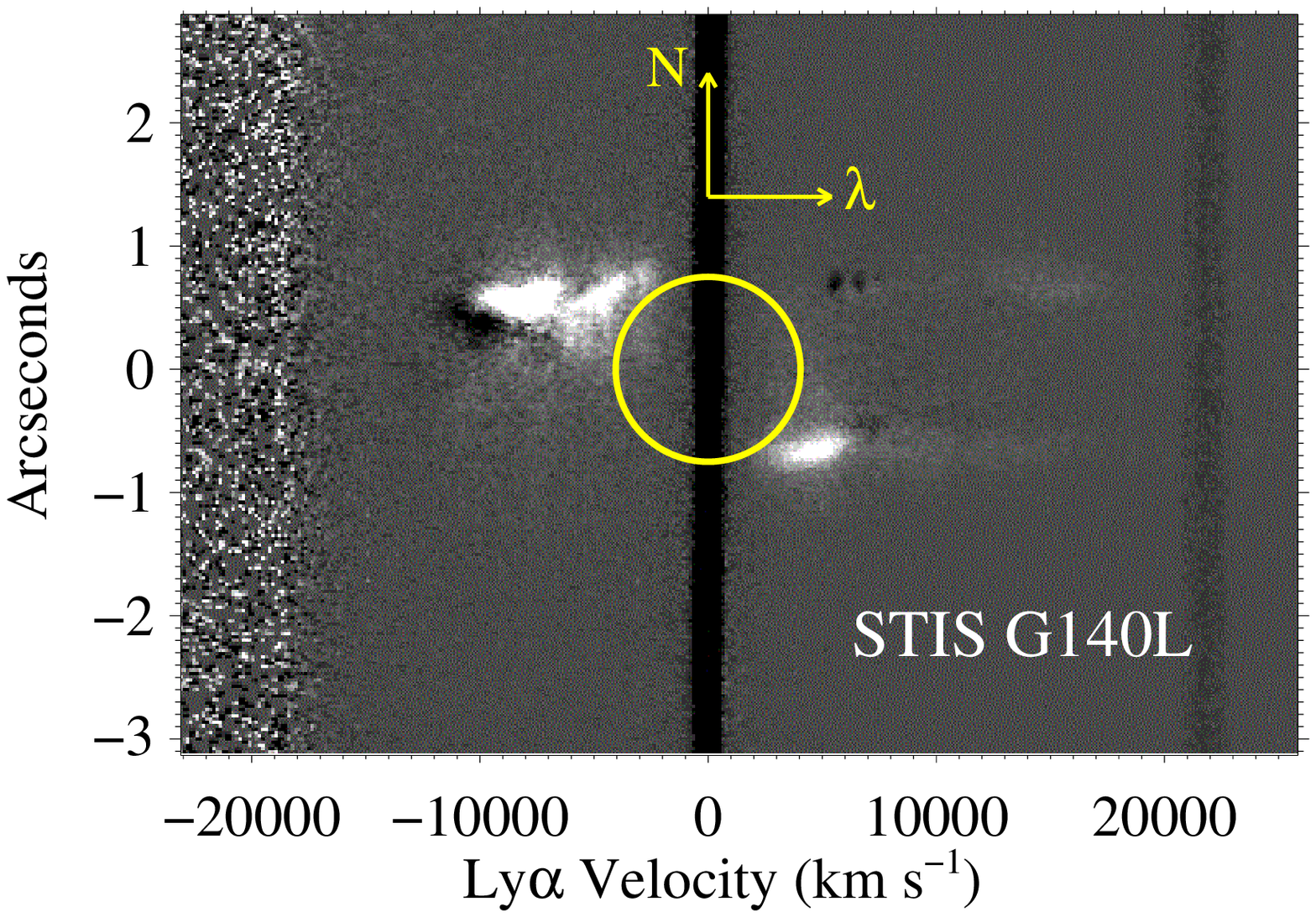,width=3.5in,angle=90}
\epsfig{figure=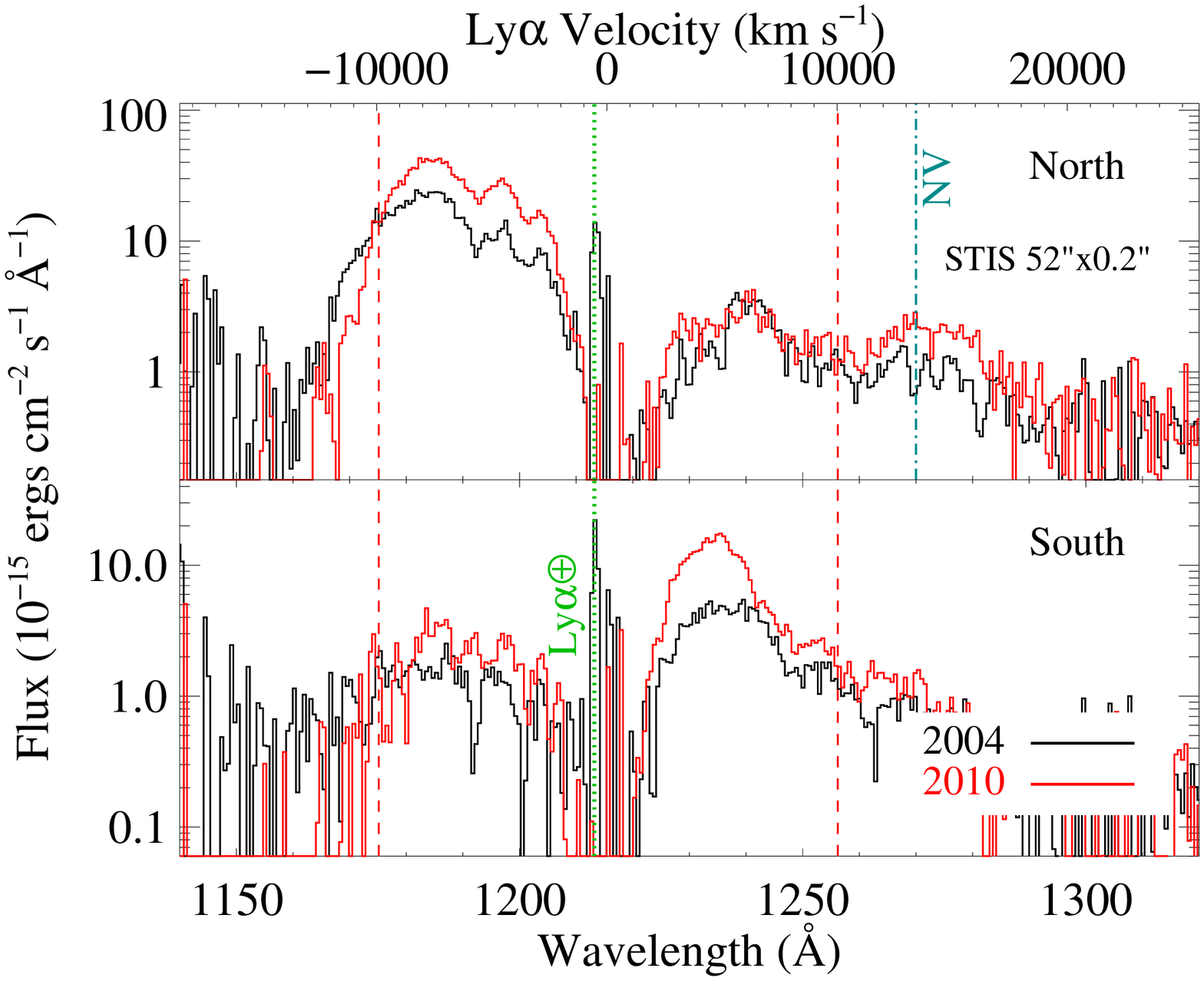,width=2.5in,angle=90}
\caption{({\it left}) Ly$\alpha$ 2010-2004 difference image, scaled such that gray indicates similar intensities in the two epochs.  The yellow circle approximates the location of the circumstellar emission ring.  ({\it right}) One-dimensional spectra of the Ly$\alpha$ emission from the North and South regions~\citep{n5}. The narrow feature labeled ``$\oplus$'' is residual emission from Earth's upper atmosphere.	
}
\end{center}
\end{figure}

The Ly$\alpha$ emission brightened from 2004 to 2010 (Fig. 5). We observe two primary features: (a) the Ly$\alpha$ emission has increased in brightness by factors of 1.6--2.4 for the North and South shock emission, and (b) the maximum Doppler shift in the northern blue shifted Ly$\alpha$ emission is decreasing as a function of time.  Note also the faint glow seen on the north and south sides at wavelengths ranging from $\sim$ 1260--1290 \AA, also visible in Figure 3. This emission cannot be attributed to Ly$\alpha$, because it would require the Ly$\alpha$ emission on the north side to be redshifted by velocities up to +20,000 km/s, while the actual Ly$\alpha$ emission on that side is blueshifted (Fig. 3).

We propose that this emission comes from fast-moving N$^{4+}$ ions in a thin layer immediately downstream from the reverse shock. Neutral or singly-ionized nitrogen atoms that cross the reverse shock and enter the shocked plasma are repeatedly ionized by collisions in the shock transition zone. As it passes through the Li-like ionization stage (N$^{4+}$), a nitrogen atom may be excited to the $2^2$P fine-structure state and emit a N~{\sc v}~1239\AA\ or 1243\AA\ photon, or it may be ionized to N$^{5+}$. The number of N~{\sc v} photons produced, per nitrogen atom passing through the reverse shock, will be equal to the ratio of the N$^{4+}$ $2^2$P excitation rate to its ionization rate. The Ly$\alpha$ excitation rate is approximately equal to the ionization rate, however the excitation rate producing the N~{\sc v} emission (which is dominated by collisions with protons and alpha particles) exceeds the ionization rate by a factor of several hundred \citep{laming96,borkowski97}. We estimate that each nitrogen atom that passes through the reverse shock will emit $\sim$~600 N~{\sc v} photons before it becomes fully ionized. Given the enriched abundance of nitrogen (N/H $\sim$~ 2 $\times$ 10$^{-4}$) 
\citep{lundqvist96}	
in the equatorial ring (and presumably in the outer debris of SN 1987A), the identification of these emissions as N~{\sc v} is plausible.  

The putative N~{\sc v} emission is redshifted on the northern side of the reverse shock because the nitrogen atoms are ionized and accelerated by the turbulent electromagnetic fields in the isotropization zone of the collisionless shock before they emit N~{\sc v} photons.  This is in contrast to the hydrogen atoms, which are not deflected significantly from free expansion when they emit Ly$\alpha$.
If our identification of this faint feature as N~{\sc v} is correct, we are seeing redshifts on the north side extending to $\sim 12,000$ km~s$^{-1}$.  The profile of the N~{\sc v} emission will be a convolution of the projected velocity distribution function of the N$^{4+}$ ions with the shock surface, both of which are unknown.  But it is probable that the line profile will also have a blueshifted wing extending to at least $\sim -12,000$ km~s$^{-1}$.  We cannot discern this wing because it will be buried under the much brighter Ly$\alpha$ emission.  A critical test of our hypothesis will be the observation of the profile of the C~{\sc iv}~$\lambda\lambda$1548,1550\AA~doublet, which we estimate to have a brightness $\sim 0.3$ times that of N~{\sc v}~$\lambda\lambda$1239,1243\AA.  The C~{\sc iv} doublet should have the same intrinsic emission profile as N~{\sc v}, but it will not be confused with Ly$\alpha$ emission and absorption.


\section{Supplementary Material}

{\bf On-line hotspot movie caption:} [ This movie displays the evolution of the
remnant of SN1987A as observed by the {\it Hubble Space Telescope}. The rapidly
fading and expanding central source is light from the inner radioactive
supernova debris. The inner circumstellar ring is glowing initially because it
was ionized by radiation from the supernova outburst. At about 1995, the first
hotspot appeared at approximately 11 o'clock. Today, the ring is entirely
encircled by hotspots. The radiation from the hotspots is caused by compression
and heating that takes place when the supernova blast wave enters fingers of
dense gas protruding inwards from the circumstellar ring.]


\begin{thebibliography}{99}
\bibitem[NOTE5(2010)]{n5} If one attributes the faint redshifted features to N~{\sc v} $\lambda\lambda$1239,1243 \AA, one should subtract 6150 km s$^{-1}$ (the Ly$\alpha$~--~N~{\sc v} velocity difference) from the Ly$\alpha$ velocity scale to measure the corresponding Doppler shift.
\bibitem[Acknowl(2010)]{acknowl10} K.F. thanks Matthew Beasley for enjoyable discussions.  
Support for this research was provided by NASA through grant GO-11181 from the Space Telescope Science Institute, which is operated by the Association of Universities for Research in Astronomy, Incorporated, under NASA contract NAS5-26555.
\end{thebibliography}
\end{document}